\documentclass[]{revtex4}

\usepackage{graphicx}
\usepackage{hyperref}

\begin{document}



\title{Light-Assisted Cold Chemical Reactions of Barium Ions with Rubidium Atoms} 
\author{Felix H.J. Hall}
\affiliation{Department of Chemistry, University of Basel, Basel, Switzerland}
\author{Mireille Aymar}
\author{Maurice Raoult}
\author{Olivier Dulieu}
\affiliation{Laboratoire Aim\'{e} Cotton, CNRS/Univ Paris-Sud/ENS Cachan, B\^{a}t. 505, Campus d'Orsay, 91405 Orsay Cedex, France}
\author{Stefan Willitsch}
 \email{stefan.willitsch@unibas.ch}
\affiliation{Department of Chemistry, University of Basel, Basel, Switzerland}



\begin{abstract}
Light-assisted reactive collisions between laser-cooled Ba$^+$ ions and Rb atoms were studied in an ion-atom hybrid trap. The reaction rate was found to strongly depend on the electronic state of the reaction partners with the largest rate constant ($7(2)\times10^{-11}$~cm$^3$~s$^{-1}$) obtained for the excited Ba$^+(6s)$+Rb$(5p)$ reaction channel. Similar to the previously studied Ca$^+$+Rb system, charge transfer and radiative association were found to be the dominant reactive processes. The generation of molecular ions by radiative association could directly be observed by their sympathetic cooling into a Coulomb crystal. Potential energy curves up to the Ba$^+(6s)$+Rb$(5p)$ asymptote and reactive-scattering cross sections for the radiative processes were calculated. The theoretical rate constant obtained for the lowest reaction channel Ba$^+(6s)$+Rb$(5s)$ is compatible with the experimental estimates obtained thus far.

\end{abstract}

\maketitle

\section{Introduction}

Following the impressive developments in the generation of translationally cold atoms, molecules and ions \cite{willitsch08b, carr09a, schnell09a, dulieu09a, willitsch12a}, the study of chemical reactions at ultra-low energies has recently emerged as a new exciting research area in chemical physics \cite{weiner99a, bell09b, dulieu11a, jin12a, henson12a, willitsch12a}. Over the past few years, ion-atom hybrid traps in which laser \cite{grier09a,zipkes10a,schmid10a,hall11a, sullivan12a} or sympathetically \cite{ravi12a,hall12a,haerter12a} cooled ions in a radio frequency (rf) ion trap are combined with ultracold atoms in a magnetic \cite{zipkes10a}, optical dipole \cite{schmid10a,haerter12a} or magneto-optical {\cite{grier09a,hall11a,hall12a,ravi12a,sullivan12a} trap have paved the way for the investigation of sub-Kelvin collisions of atomic and since recently also molecular \cite{hall12a} ions with ultracold atoms. At these collision energies, the quantum character of the collision can influence elastic \cite{cote00a} as well as inelastic \cite{krych11a,belyaev12a} scattering cross sections and the validity of ab-initio and quantum scattering calculations as well as reactive capture models can be checked accurately \cite{grier09a, hall11a, rellergert11a, hall12a, sullivan12a, hall13a}. Moreover, the possibility to introduce near-resonant laser fields reveals the prominence of radiation-driven chemical processes and highlights the role of electronically excited states in cold collisions through the determination of state-specific rate constants \cite{hall11a, rellergert11a, ratschbacher12a,sullivan12a, hall13a}.

In the present paper, we report on a joint experimental and theoretical study of light-assisted reactive collisions between laser-cooled Ba$^+$ ions and Rb atoms extending the data on collisions between the ground-state species reported by Schmid et al. \cite{schmid10a}. Utilising hybrid-trapping technology average collision energies $\langle E_{\text{coll}}/k_B\rangle \geq 600$~mK were achieved. Rate constants for excited reaction channels populated by near resonant laser fields  were determined. We find significant enhancement of the rates in electronically excited channels which is rationalised with reference to computed potential energy curves and transition dipole moments. Quantum-scattering calculations predict that reactive collisions between the ground-state collision partners predominantly lead to the radiative association (RA) of molecular ions which is supported by the experimental observation of BaRb$^+$ ion formation. The collision energy dependence of the reactive processes is studied and the experimental rate constants are compared with theoretical predictions. Finally, the present results on Ba$^+$+Rb are compared with and contrasted to the findings in other ion-atom hybrid systems. 

\section{Methods}
\label{methods}

The experimental setup used for the simultaneous trapping of atomic ions and neutral atoms has already been described previously \cite{hall11a,hall13a, willitsch12a} and only aspects relevant to the present study are described here. 

\begin{figure}
\begin{center}
\includegraphics[width=\textwidth]{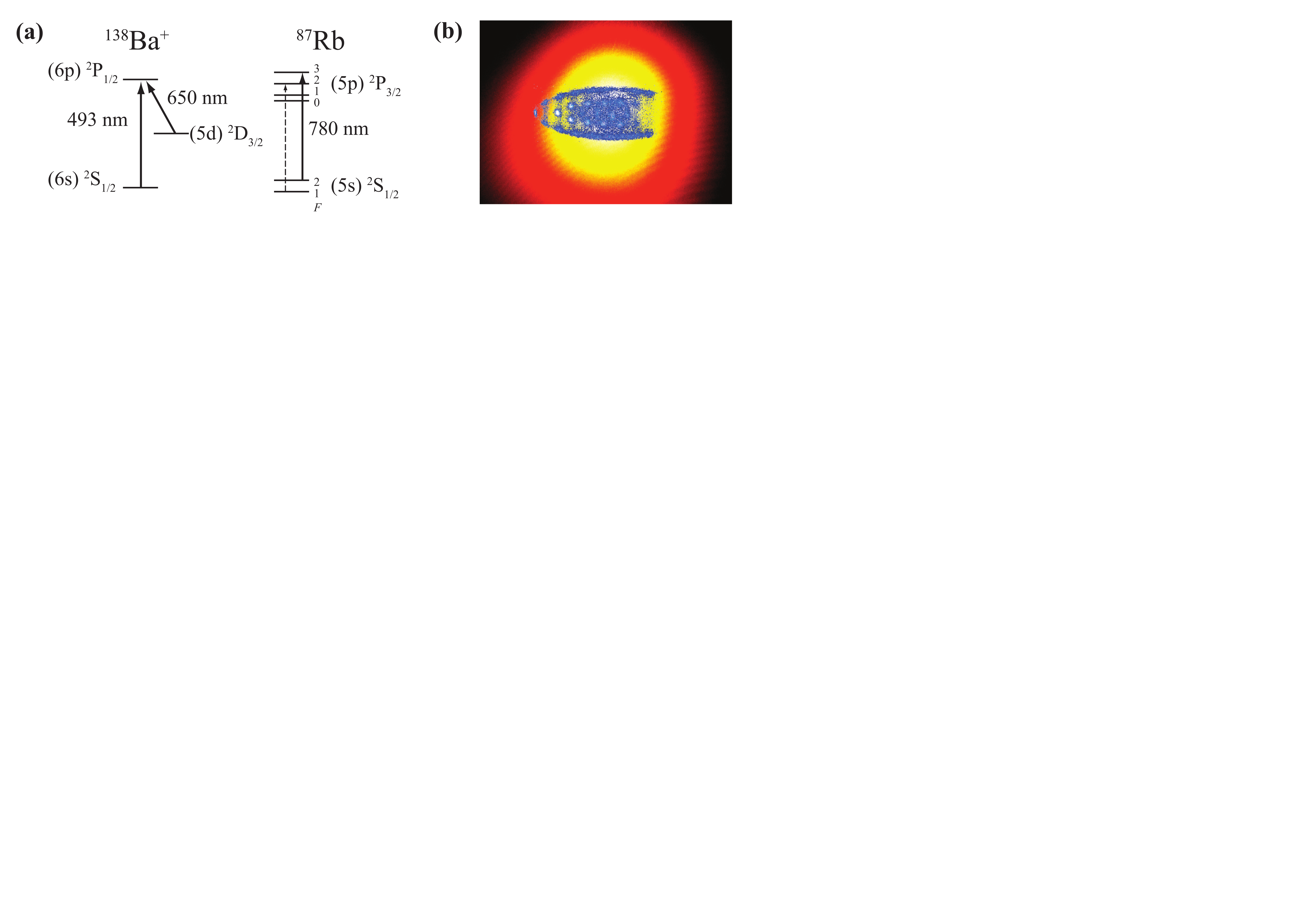}%
\caption{\label{elevel} (a) Laser cooling schemes for $^{138}$Ba$^+$ and $^{87}$Rb. (b) Superposed false-colour fluorescence images of a Coulomb crystal of $^{138}$Ba$^+$ ions (blue) and a cloud of ultracold $^{87}$Rb atoms (yellow-red) in the hybrid trap. The asymmetric shape of the $^{138}$Ba$^+$ Coulomb crystal is caused by non-fluorescing sympathetically-cooled Ba$^+$ isotopes, see text for details.}%
\end{center}
\end{figure}

Ba$^+$ ions were produced by non-resonant photoionisation of a beam of Ba atoms with their natural isotopic abundances at the centre of a linear radio frequency (RF) ion trap \cite{hall11a, willitsch12a, hall13a}. Trapped $^{138}$Ba$^+$ ions were laser cooled to secular temperatures of $\approx$ 10~mK using a laser beam at a wavelength around 493~nm slightly red detuned from the $(6s)~^{2}S_{1/2} \rightarrow (6p)~^{2}P_{1/2}$ transition. A repumper laser at 650~nm was used to repump population on the $(5d)~ ^{2}D_{3/2} \rightarrow (6p)~ ^{2}P_{1/2}$ transition (see figure 1(a)). Under these conditions, the Ba$^+$ ions formed Coulomb crystals (Fig.~\ref{elevel}(b)) \cite{willitsch12a} which were imaged by collecting the spatially resolved laser-cooling fluorescence of the ions using a charge coupled device (CCD) camera coupled to a microscope. 

$^{87}$Rb atoms were produced by a getter source and laser cooled and trapped in a magneto-optical trap (MOT) \cite{hall11a, hall13a}. 95\% of the MOT laser beam intensity was detuned by $\approx 20$~MHz from the $(5s)~^{2}S_{1/2}, F = 2 \rightarrow (5p)~^{2}P_{3/2}, F = 3$ cooling transition of $^{87}$Rb around 780~nm. The remaining 5\% were tuned into resonance with the $(5s)~^{2}S_{1/2}, F = 1 \rightarrow (5p)~^{2}P_{3/2}, F = 2$ repumper transition (see Fig.~\ref{elevel}(a)). The number density ($n_{Rb} \approx 1\times10^9~$cm$^{-3}$) and temperature ($T$ = 90 $-$ 150 $\mu$K) of the atoms in the MOT were determined using standard fluorescence measurement and time-of-flight methods, respectively \cite{shah07a, pradhan08a}. To avoid multi-photon ionisation of Rb atoms in the excited $(5p)~^2P_{3/2}$ state by the Ba$^+$ cooling laser, the 493~nm and 780~nm laser beams were alternately blocked using a mechanical chopper operating at a frequency of 1000~Hz. 

The populations of the $(5s)~^2S_{1/2}$ and $(5p)~^2P_{3/2}$ levels of the $^{87}$Rb atoms in the MOT were determined from the fluorescence intensity of the atom cloud using the steady-state solution of the optical Bloch equations (OBE) for a two level system \cite{shah07a}. The steady-state populations of the $(6s)~^2S_{1/2}, (6p)~^2P_{3/2}$, and $(5d)~^2D_{3/2}$ levels of the Coulomb-crystallised $^{138}$Ba$^+$ ions were calculated using an 8-level optical Bloch equation model including the effects of magnetic fields \cite{gingell10a, oberst99a}. This method has been tested and calibrated previously by comparisons with the fluorescence yield of single Ca$^+$ ions \cite{hall13a}.

Because of the large depth of the ion trap ($\approx 4$~eV), the ionic reaction products remained trapped and were sympathetically cooled into the Coulomb crystal by their interaction with the remaining laser-cooled Ba$^+$ ions. The masses of the different ion species were determined by resonance-excitation mass spectrometry (REMS) as described in Refs. \cite{hall11a, hall12a}. Molecular-dynamics (MD) methods were used to simulate the laser-cooling fluorescence images of the ions in order to determine the composition and kinetic-energy characteristics of the Coulomb crystals \cite{bell09a, willitsch12a}. 

The BaRb$^+$ complex has the same valence electronic structure as alkali dimers with two external electrons moving in the field of the Rb$^+$ and Ba$^{2+}$ ionic cores. Thus, the electronic potential curves (PECs) and transition dipole moments (TDMs) of the BaRb$^+$ molecule can be conveniently calculated using a full configuration interaction approach performed in a configuration space generated by the large Gaussian basis sets reported in Refs.\cite{aymar05a, aymar06a, aymar12a}. The two-electron Hamiltonian involves an effective core potential for the Ba$^{2+}$ \cite{fuentealba85a, fuentealba87a} and for the Rb$^+$ \cite{aymar05a} ions together with core-polarization potentials \cite{mueller84a, guerout10a} to account for the correlation between core and valence electrons. The spin-orbit correction to the PECs has not been considered in the present work.

The cross sections for RA and radiative charge transfer (RCT) are computed using the same method as in our previous study on Ca$^+$+Rb reactive collisions \cite{hall13a}. The electronic TDM matrix elements between the initial state (a ground-state Ba$^+$ ion colliding with a ground-state Rb atom) and the final states (either a ground-state BaRb$^+$ molecule in the case of RA or a pair of free Ba and Rb$^+$ particles for RCT) are integrated over the internuclear distance $R$ using the Milne phase amplitude method \cite{milne30a, korsch77a}. Because of the large mass of the collisional complex, a large number of partial waves contribute to the collisions even at the low energies achieved in the present study, so that this approach allows for an efficient localization of orbiting resonances (see below).

\section{Results and discussion}

\subsection{Reaction channels and channel-specific rate constants}

\begin{figure}
\begin{center}
\includegraphics[width=\textwidth]{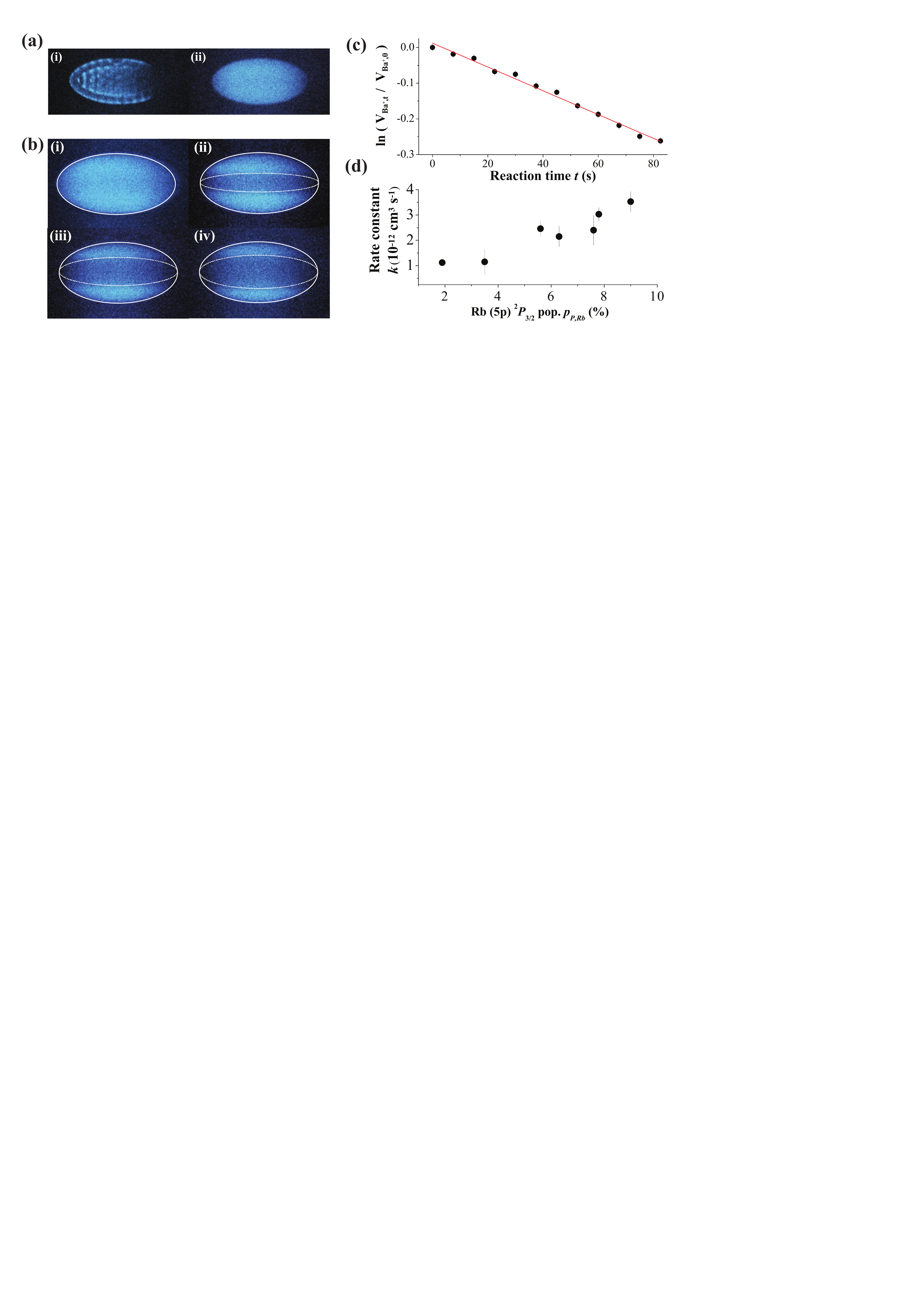}%
\caption{\label{isomix} (a) Demonstration of Ba$^+$ isotope mixing in the Coulomb crystals: (i) Uni-directional laser cooling results in radiation pressure on the $^{138}$Ba$^+$  ions and a localization of the non-fluorescing Ba$^+$ isotopes (predominantly $^{137}$Ba$^+$, $^{136}$Ba$^+$, $^{135}$Ba$^+$ and $^{134}$Ba$^+$) on the right-hand side of the crystal. (ii) The radiation pressure is cancelled in a bi-directional laser-cooling configuration leading to a complete mixing of the isotopes in the crystal. (b) Images of a crystal of Ba$^+$ ions over the course of a typical reaction with Rb atoms. Solid white ellipses highlight the approximative boundaries of the crystal of fluorescing Ba$^+$ ions, and dashed white ellipses approximately delimit the dark central core of product Rb$^+$ ions. (c) Plot of $\ln(V_{Ba^+,t} / V_{Ba^+,0})$ against reaction time following pseudo-first-order kinetics. The red line represents a linear regression to the data. See text for details. (d) Effective rate constant $k$ as a function of the population in the excited Rb$(5p)~^2P_{3/2}$ state.}%
\end{center}
\end{figure}

Fig. \ref{isomix} (a) (i) shows a typical Coulomb crystal of laser-cooled Ba$^+$ ions used in the present experiments. Only the $^{138}$Ba$^+$ isotope (72\% natural abundance) is laser-cooled. The other isotopes (predominantly $^{137}$Ba$^+$ (11\%), $^{136}$Ba$^+$ (8\%), $^{135}$Ba$^+$ (7\%) and $^{134}$Ba$^+$(2\%)) are sympathetically cooled and appear as a dark non-fluorescing region on the right-hand side of the crystal. The $^{138}$Ba$^+$ ions are localized on the left-hand side because of radiation-pressure forces exerted by the cooling laser beam propagating along the trap symmetry axis from right to left in the image. When the cooling laser is retro-reflected after leaving the trap, the radiation pressure is cancelled and a complete mixing of all Ba$^+$ isotopes in the crystal can be observed (Fig. \ref{isomix} (a) (ii)).

As reactions of the Ba$^+$ ions with Rb proceeded, the formation of a dark core about the central axis of the crystals could be observed which consists of sympathetically-cooled product ions (Fig.~\ref{isomix}(b)(ii-iv)). In principle, all Ba$^+$ isotopes can participate in the reactions. However, the laser-cooled $^{138}$Ba$^+$ species reacts preferentially as a consequence of the electronic excitation (see below).

At the low number densities of Rb atoms in our MOT ($n_{Rb} \approx 1\times10^9~$cm$^{-3}$), ternary reactive processes which are observed in hybrid-trap experiments using denser atom clouds \cite{haerter12a} are precluded and the reaction kinetics follow a second-order rate law. The constant replenishment of the Rb atoms in the MOT from background vapour results in pseudo-first-order kinetics defined by an integrated rate law of the form $\ln(V_{Ba^+,t}/V_{Ba^+,t=0}) = -k_{pfo}t$ \cite{willitsch08a, hall11a}. $k_{pfo}$ denotes the pseudo-first order rate constant, $t$ is the reaction time, and $V_{Ba^+,t=0}$ and $V_{Ba^+,t}$ are the volumes occupied by the reactant ions at $t = 0$ and $t$, respectively. The second order rate constant $k$ is obtained via $k = k_{pfo}/n_{Rb}$.

Reaction rate constants were determined by monitoring the decrease of the volume occupied by Ba$^+$ ions as a function of the reaction time. For this purpose, the volume $V_{\text{tot}}$ of the entire crystal (i.e., reactant and product ions) was estimated by fitting an ellipsoid around its perimeter and subtracting the volume $V_{\text{core}}$ of the central core which consists only of sympathetically-cooled product ions (see Figs.~\ref{isomix}(b)(ii-iv)). The core of sympathetically-cooled ions exhibits in general a more complex shape \cite{hornekaer01a}. However, in order to enable a quick and efficient analysis of the large number of images accumulated in the experiments, $V_{\text{core}}$ was approximated by fitting another ellipsoid to the edges of the dark core as indicated in Figs.~\ref{isomix}(b)(ii-iv). A plot of $\ln(V_{Ba^+,t}/V_{Ba^+,t=0})$ against $t$ is presented in Fig.~\ref{isomix}(c). The data almost perfectly conform to a pseudo-first-order behaviour justifying our approximations for the calculation of the crystal volumes. From a linear regression of the data, $k$ was determined to be $2.0(5)\times10^{-12}$ cm$^3$ s$^{-1}$ for the experiment shown.

By inspection of the energy level diagram Fig.~\ref{elevel}(b), it can be expected that the following channels contribute to the observed reactions: (i) Ba$^+$(6s)+Rb(5s), (ii) Ba$^+$(5d)+Rb(5s), (iii) Ba$^+$(6p)+Rb(5s) and (iv) Ba$^+$(6s)+Rb(5p). Because of the alternate blocking of the Ba$^+$ and Rb cooling laser beams (see section \ref{methods}), reactions with simultaneously excited collision partners were precluded. The rate constants for the processes (i)-(iv) are designated $k_s,  k_d,  k_p,$ and $k^*_s$. The experimentally observed rate constant $k$ is a weighted average over the contributions from all possible channels and can be formulated as \cite{hall11a}:
\begin{equation}
k = \frac{1}{2}[(p_{S,Ca^+} + p_{S,Rb})k_s + p_{P,Ca^+}k_p + p_{D,Ca^+}k_d + p_{P,Rb}k^*_s], \label{rateeq}
\end{equation}
where $p_{i,j}$ are the steady state populations of the electronic state $i$ of the reaction partner $j$. We observe a strong dependence of $k$ on the populations in the Ba$^+(6p)$ and Rb$(5p)$ states (see Fig. \ref{isomix} (d)). To quantify these dependencies, $k$ was measured as a function of the level populations for a set of 19 different laser intensities and detunings from which the channel-specific rate constants $k_i$ were determined in a multi-dimensional least-squares fit of the data to Eq.~(\ref{rateeq}). Some $k_i$ were found to be consistent with zero in the fit so that their value was associated with an upper bound given by their $1\sigma$ uncertainty. The fit yielded $k_s \leq 5\times10^{-13}$~cm$^3$~s$^{-1}$, $k_p = 2(1)\times10^{-11}$~cm$^3$~s$^{-1}$, $k_d \leq 1\times10^{-12}$~cm$^3$~s$^{-1}$, $k^*_s = 7(2)\times10^{-11}$~cm$^3$~s$^{-1}$. The numbers in parentheses denote the $1\sigma$ statistical uncertainty of the fitted values. The rate constants are affected by an additional systematic error of $\approx 50$\% resulting from the uncertainty in the determination of the Rb number density in the MOT. 

\subsection{Reaction products}

\begin{figure}
\begin{center}
\includegraphics[width=\textwidth]{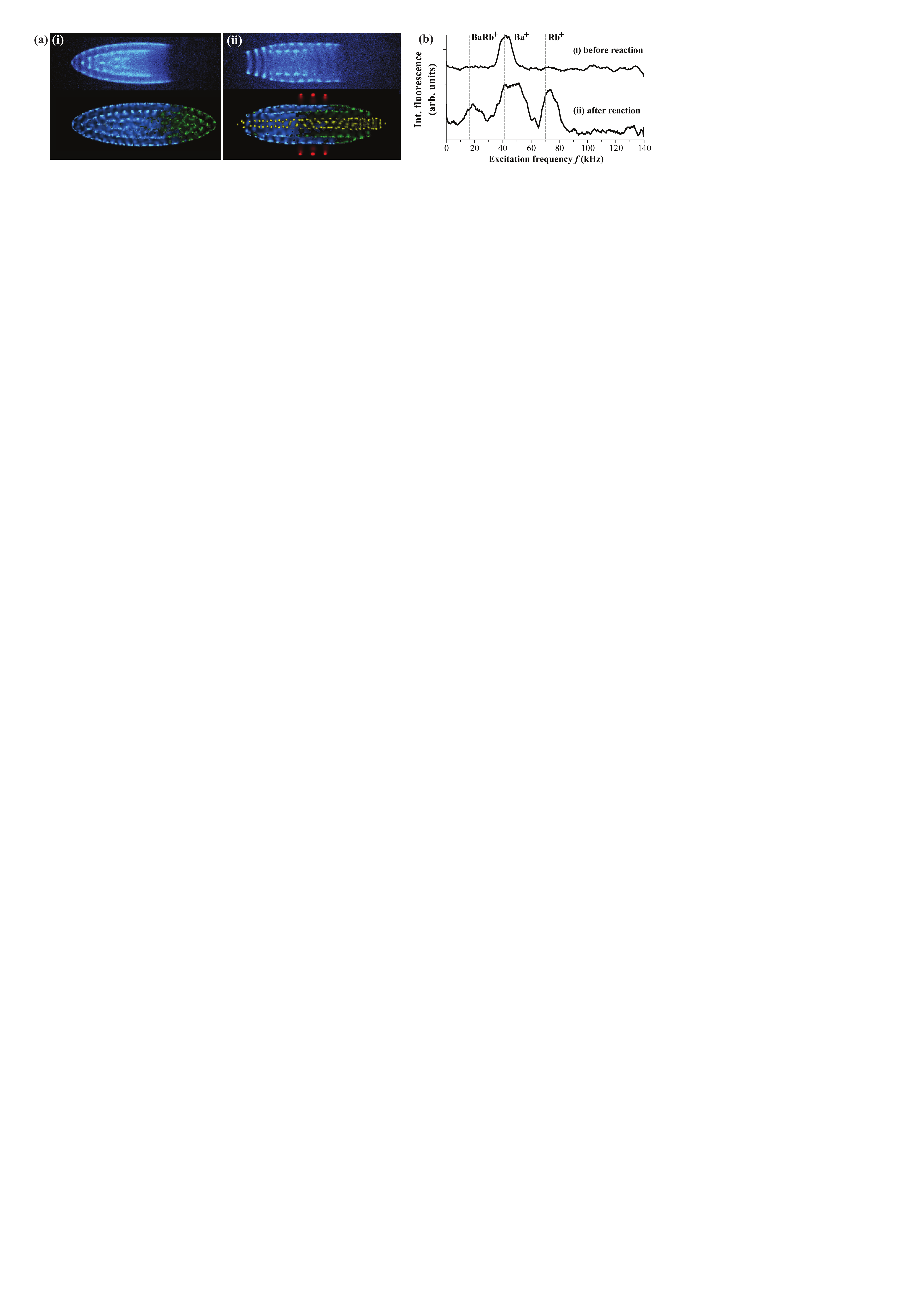}%
\caption{\label{scans} (a) Experimental false-colour fluorescence images of Ba$^+$ Coulomb crystals (upper panels) and their molecular-dynamics (MD) simulations (lower panels) (i) before and (ii) after reaction with ultracold Rb atoms. Sympathetically-cooled ions have been made visible in the MD simulations for clarity. Colour code: blue: $^{138}$Ba$^+$, green: lighter Ba$^+$ isotopes, yellow: $^{87}$Rb, red: BaRb$^+$ molecular ions. (b) Resonance excitation mass spectra (REMS) of crystals (i) before and (ii) after reaction. See text for details.}%
\end{center}
\end{figure}

Fig. \ref{scans} (a) shows images of a representative Coulomb crystal (i) before and (ii) after reaction together with their MD simulations. In these experiments, a uni-directional laser-cooling configuration of the Ba$^+$ ions was adopted to ensure a better localization of the ions facilitating the analysis of the crystal composition. 

Fig.~\ref{scans}(b) shows representative resonant-excitation mass spectra of crystals recorded (i) before and (ii) after reaction. Whereas only one feature corresponding to the Ba$^+$ reactant ions was observed before reaction, the spectrum after reaction shows three peaks assigned to Rb$^+$, Ba$^+$, and BaRb$^+$ motional resonances. The dashed lines in Fig.~\ref{scans} (b) indicate the theoretical single-ion radial excitation frequencies \cite{willitsch12a} of the different ion species. 

In the MD simulations in Fig.~\ref{scans}(a), the dark Ba$^+$ isotopes as well as the product ions have been made visible to indicate their spatial distribution. The best reproduction of the experimental fluorescence image in (ii) was obtained by assuming a composition of 170 $^{138}$Ba$^+$ (blue), 116 lighter isotopes of Ba$^+$ in their relative natural abundances (green), 90 $^{87}$Rb$^+$ (yellow), and 40 BaRb$^+$ (red). From the relative product ion yields, the BaRb$^+$ : Rb$^+$ branching ratio (representing an average over all reaction channels) was estimated to be $\approx$1 : 2.25.

In our previous study on light-assisted Ca$^+$+Rb collisions \cite{hall11a, hall13a}, the formation of Rb$_2^+$ ions was observed. Their occurrence was attributed to consecutive reactions of CaRb$^+$ with Rb. Another mechanism leading to their generation could be RA of the Rb$^+$ product ions with Rb atoms from the MOT. In the present case, however, no clear feature assignable to Rb$_2^+$ (which would be located in between the broad BaRb$^+$ and Ba$^+$ resonances in Fig. \ref{scans} (b) (ii)) was observed in the mass spectra. We therefore conclude that the formation of Rb$_2^+$ constituted at best only a minor process in the present experiments and was therefore neglected in the modeling of the Coulomb-crystal composition after reaction.

\subsection{Potential energy curves and reaction mechanisms}

Fig. \ref{pec} shows the computed non-relativistic potential energy curves of the lowest electronic states of BaRb$^+$ up to the Ba$^+(6s)$+Rb$(5p)$ asymptote. The PECs for the lowest electronic states exhibit an overall qualitative agreement with those reported in Refs. \cite{knecht10a, krych11a}. In particular, a double-minimum structure for the $A~^1\Sigma^+$ PEC which arises from an avoided crossing with the $C~^1\Sigma^+$ state has also been found in the calculations of Refs. \cite{knecht10a, krych11a}. In the results of Knecht et al. \cite{knecht10a}, the top of the barrier appears to be located slightly above the dissociation limit of this state both in their non-relativistic and relativistic calculations, whereas in the current work the barrier height is predicted to be slightly smaller than the dissociation energy. Krych et al. \cite{krych11a} demonstrated that the height of the barrier dramatically depends on the size of the configuration space used for the calculations. As discussed below, this feature has a crucial influence on the RA rate in the lowest reaction channel. 

The entrance channels accessible in the present experiments are indicated in bold. The lowest channel Ba$^+(6s)$+Rb$(5s)$ correlates with the $A~^1\Sigma^+$ and $a~^3\Sigma^+$ excited electronic states of the BaRb$^+$ molecule. From the $A~^1\Sigma^+$ state, dipole-allowed radiative processes connecting to the $X~^1\Sigma^+$ ground electronic state lead to RCT forming Rb$^+$ ions and Ba atoms and RA of BaRb$^+$ molecular ions. 

\begin{figure}[!t]
\begin{center}
\includegraphics[width=0.7\textwidth]{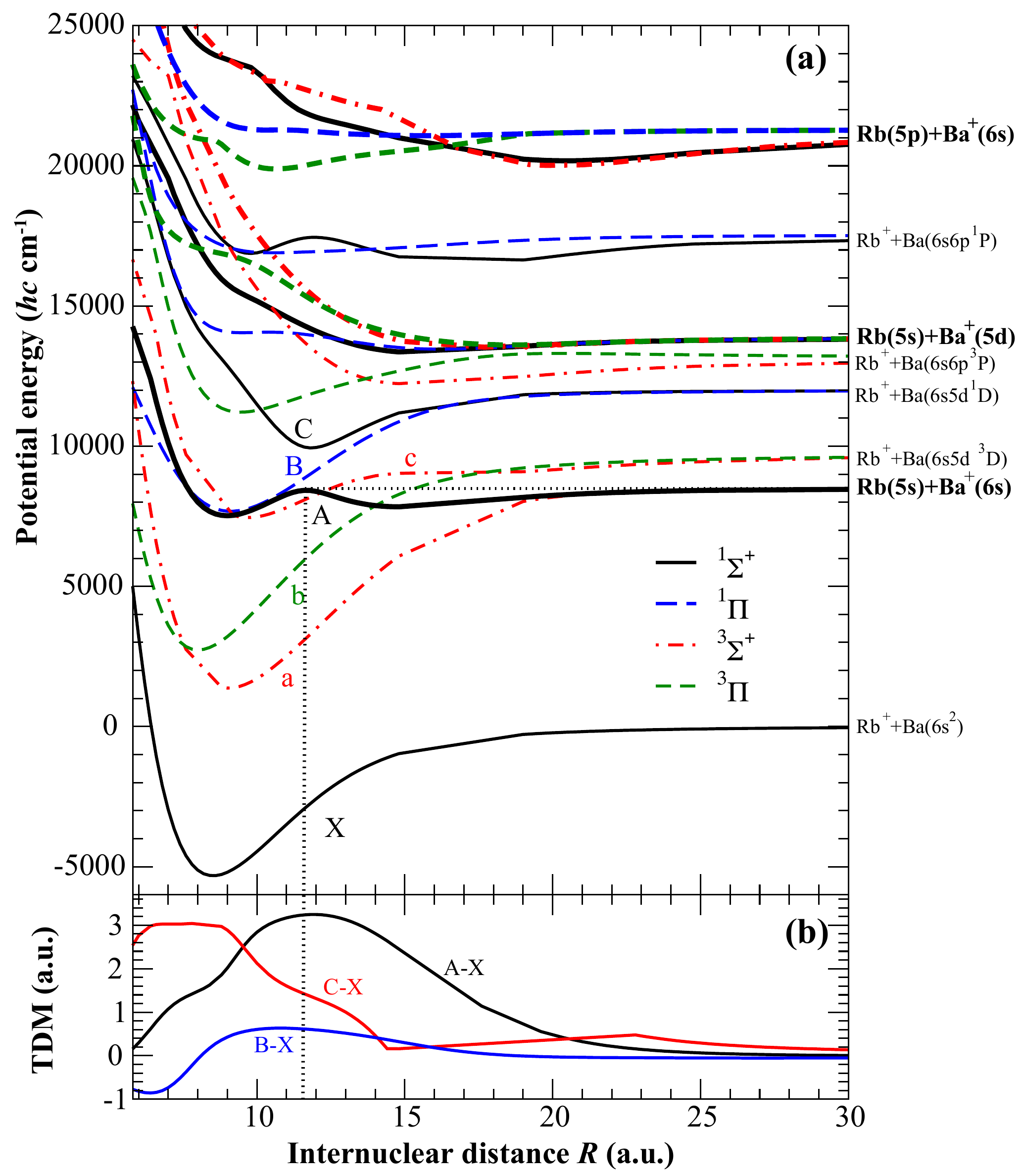}%
\caption{\label{pec} (a) Computed non-relativistic potential energy curves (PECs) of the BaRb$^+$ molecule. The lowest states have been labeled using standard spectroscopic notation. Collision channels relevant for the present study have been indicated in bold face. States with $\Delta$ symmetry are not shown. (b) Computed transition dipole moments (TDMs) between the ground ($X$) and the two lowest $^1\Sigma^+$ states ($A$ and $C$) as well as the lowest $^1\Pi$ state ($B$). The horizontal dotted line represents the energy of the initial collisional state in the lowest channel, while the vertical dotted line locates the position of the top of the barrier in the $A$ state.}%
\end{center}
\end{figure}

Non-adiabatic transitions induced by the spin-orbit interaction around the curve crossings with the $b, B$ and $c$ states visible in Fig. \ref{pec} could in principle lead to non-radiative charge transfer (NRCT). These states, however, correlate with Ba$(6s5d)$+Rb$^+$ asymptotes which are energetically not accessible at the low collision energies in the present experiments so that the NRCT channels are closed. This situation contrasts with the dynamics of the lowest reaction channel in the Ca$^+$+Rb system in which the energetic order of the analoguous asymptotes is reversed and NRCT was found to be the dominant process \cite{hall11a, hall13a, tacconi11a}. 

The low reaction rate constant $k_d \leq 1\times10^{-12}$~cm$^3$~s$^{-1}$ observed for the Ba$^+(5d)$+Rb$(5s)$ can qualitatively be explained by the repulsive character of the relevant PECs. The Ba$^+(6s)$+Rb$(5p)$ channel exhibits the largest rate constant among the channels observed ($k^*_s = 7(2)\times10^{-11}$~cm$^3$~s$^{-1}$). The fast rate in this channel could be caused by radiative processes connecting to the manifold of lower-lying electronic states. Moreover, the collision rate is expected to be about 4-5 times larger than in the other channels because of the strongly attractive interaction between the charge of Ba$^+$ and the permanent quadrupole moment of Rb in the $^2P_{3/2}$ state \cite{hall12a}.

We note that the mechanism for the suppression of reaction channels involving excited neutral atoms discussed in Ref. \cite{sullivan12a} does not seem to play an important role under the conditions of the present study. In the present case, the Condon point for excitation to the Ba$^+(6s)$+Rb$(5p)$ curve is located at an internuclear distance $R\approx 1000$~a$_0$. At smaller distances, laser excitation of the atom is suppressed as a consequence of the shift of the electronic resonance induced by the interaction with the ion. However, at collision energies $E_{\text{coll}}/k_{\text{B}}\approx1$~K typical for the present study, the collision partners can approach over distances exceeding 9000~a$_0$ during the lifetime of the excited Rb$(5p)~^2P_{3/2}$ state so that reactions with laser-excited Rb atoms can readily occur. Similar observations have also been made in N$_2^+$+Rb collisions at energies down to $E_{\text{coll}}/k_{\text{B}}=20$~mK \cite{hall12a}.

Similar to Ca$^+$+Rb \cite{hall11a,hall13a}, a large rate constant ($k_p = 2(1)\times10^{-11}$~cm$^3$~s$^{-1}$) was also obtained for the reaction channel with the highest electronic excitation of the ion accessible by the present laser-cooling scheme, i.e.,  Ba$^+(6p)$+Rb$(5s)$. The PECs of the highly-excited electronic states correlated with the  Ba$^+(6p)$+Rb$(5s)$ asymptote could not be reliably extracted from the present set of calculations because their pattern was found to be too complex to follow them adiabatically from large to small internuclear distances. Further investigations of these states are currently under progress. However, the electronic structure at high excitation energies is qualitatively similar to the one found in other mixed alkaline-earth-ion alkali-atom systems such as CaRb$^+$ \cite{hall11a}. At high excitation energies, CaRb$^+$ exhibits a high density of electronic states correlating asymptotically with excited electronic states of neutral Ca. The high density of states promotes opportunities for RA and RCT as well as NRCT mediated by a multitude of curve crossings. Apart from CaRb$^+$ \cite{hall11a}, similar effects have recently been invoked to explain the dynamics of light-assisted processes in the BaCa$^+$ system \cite{sullivan12a}. The high density of electronic states occurring at higher excitation energies is a common feature of mixed alkaline-earth-ion alkali-atom systems \cite{dulieu13a} and it appears that the fast kinetics which has been observed in electronically excited channels across several systems \cite{hall11a, ratschbacher12a, sullivan12a} is a direct consequence of the associated radiative and non-adiabatic dynamics.

\subsection{Radiative cross sections and collision-energy dependence}
\label{rcs}

In the present experiments, the collision energy was dominated by the micromotion of the ions, i.e., their fast motion driven by the RF fields \cite{willitsch12a, hall12a}. By contrast, the contribution of the Rb atoms was negligible on account of their low temperature ($\leq$ 150~$\mu$K). The micromotion leads to non-Maxwellian ion velocity distributions which were characterised using MD simulations \cite{bell09a}. As the micromotion energy depends on the position of the ions in the trap \cite{willitsch12a}, the average collision energies $\langle E_{coll}/k_B\rangle$ in the experiments were varied by changing the size and shape of the Coulomb crystals \cite{bell09a, hall13a}. Fig.~\ref{results}(a) shows experimental (upper panels) and simulated (lower panels) images of Coulomb crystals with their collision energy distributions given in Fig.~\ref{results}(b). Fig. \ref{results} (c) shows the experimental (channel-averaged) rate constants $k$ for the crystals in Fig. \ref{results} (a) plotted against the average collision energy calculated from the distributions in Fig. \ref{results} (b). The rate constant was found to be nearly invariant with average collision energy within the uncertainty limits, with a slight increase observable at the highest energies. A similarly flat dependence of $k$ on $E_{\text{coll}}$ was also observed in the Ca$^+$ + Rb system \cite{hall11a, hall13a}. 

\begin{figure}
\begin{center}
\includegraphics[width=\textwidth]{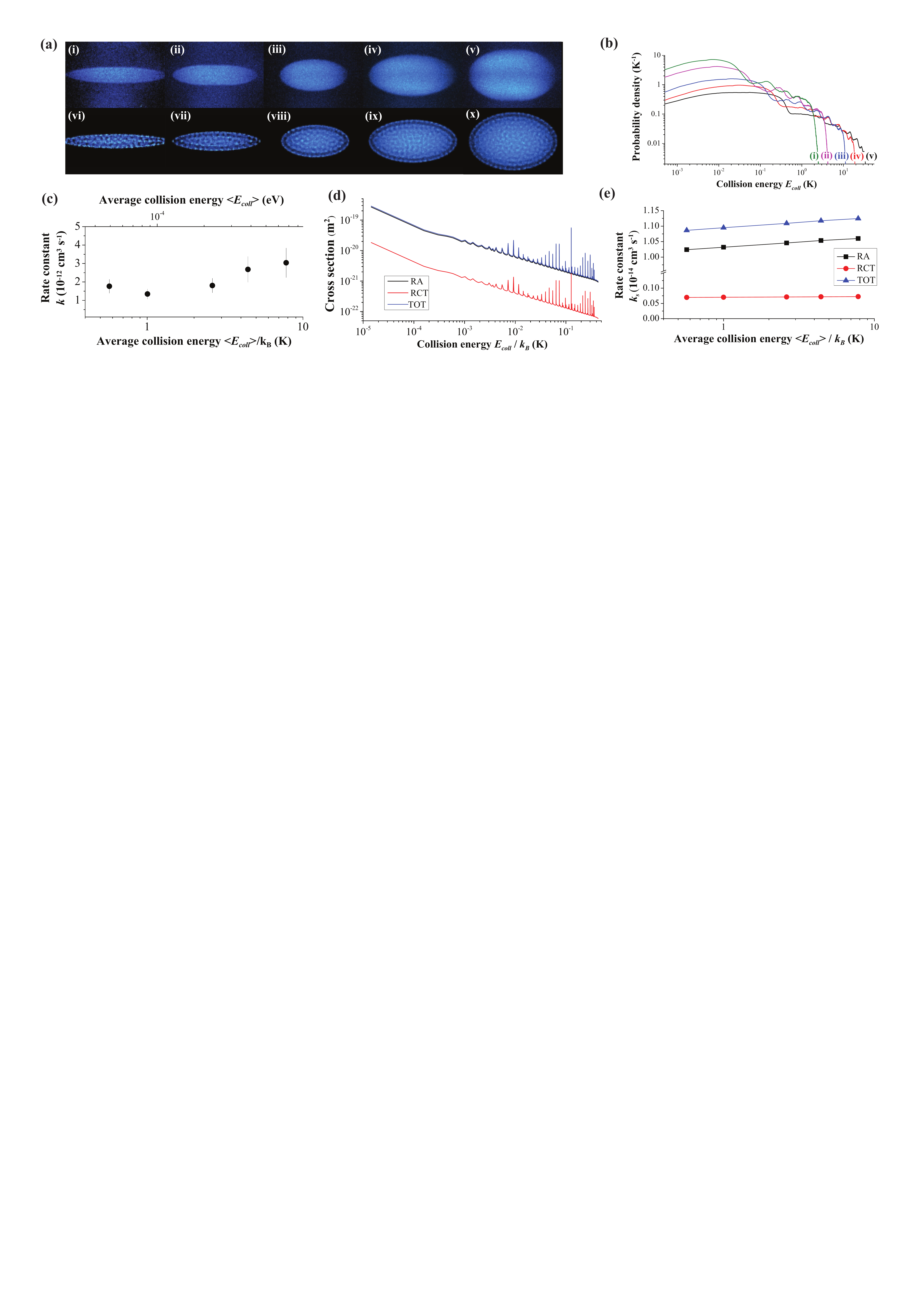}%
\caption{\label{results} (a) False-colour fluorescence images of Coulomb crystals of Ba$^+$ ions of varying shapes and sizes ((i)-(v)) and their MD simulations ((vi)-(x)). (b) Energy distributions for collisions with ultracold Rb atoms of the crystals shown in (a). (c) Dependence of the state-averaged rate constant $k$ on the average collision energy $\langle E_{\text{coll}}\rangle$. Error bars denote the $2\sigma$ statistical uncertainty of at least three consecutive measurements. (d) Theoretical cross sections for radiative association (RA) and radiative charge transfer (RCT) for the lowest entrance channel Ba$^+(6s)$+Rb$(5s)$. (e) Theoretical rate constant $k_s$ for the lowest channel Ba$^+(6s)$+Rb$(5s)$ obtained by velocity-averaging the total radiative cross section over the relative velocity distributions. See text for details.}%
\end{center}
\end{figure}

Fig. \ref{results} (d) shows the theoretical cross sections $\sigma$ for RA (black) and RCT (red) in the Ba$^+(6s)$+Rb$(5s)$ entrance channel. The cross sections for RA from the $A~^1\Sigma^+$ to the $X~^1\Sigma^+$ state are predicted to be about an order of magnitude larger than for RCT on account of the more favourable Franck-Condon factors for transitions to bound vibrational levels of the $X$ state arising from the double-well structure of the $A~^1\Sigma^+$ PEC. By comparison, the non-relativistic $^1\Sigma^+$ PEC associated with the lowest entrance channel in Ca$^+$+Rb collisions exhibits no such feature and the relative efficiencies of RCT and RA are reversed \cite{hall13a}. The cross sections exhibit pronounced modulations with collision energy identified as shape resonances arising from the trapping of the collisional wave function behind the centrifugal barrier at well-defined collision energies. The predicted values for the total radiative cross section in the energy interval between 1 and 100~mK ($\sigma\approx 10^{-19}-10^{-20}$~m$^{-2}$) are slightly lower than the experimental estimate from Ref. \cite{schmid10a} ($\sigma\approx 10^{-18}-10^{-19}$~m$^{-2}$).

In Fig. \ref{results} (e), theoretical rate constants $k_s$ for the lowest entrance channel Ba$^+(6s)$+Rb$(5s)$ are shown which were obtained by averaging the total radiative (i.e., RA+RCT) cross sections over the velocity distributions of the Coulomb crystals. Because of convergence problems arising from the double-minimum structure of the $A$ state PEC, it was not possible to calculate radiative cross sections for high-angular-momentum collisions arising at energies $E_{\text{coll}}/k_{\text{B}}>400$~mK. However, calculations in similar systems have shown that the radiative cross sections show a smooth and monotonous energy dependence at low energies \cite{hall13a}. Thus, in order to obtain a theoretical estimate of the rate constants, the base line of the total cross section curve (black in Fig. \ref{results} (d)) was extrapolated over the entire energy range relevant for the present study ($E_{\text{coll}}\leq40$~K). As shown in Ref. \cite{hall13a}, the narrow shape resonances only insignificantly contribute to the integrated cross section and could therefore be neglected. The rate constant $k_s$ for the lowest channel was then calculated at the average collision energies corresponding to the experimental data points in Fig. \ref{results} (c) by velocity-averaging the extrapolated cross sections over the relative velocity distributions derived from the collision-energy distributions in Fig. \ref{results} (b).  As can be seen from Fig. \ref{results} (e), the predicted rate constant $k_s$ slightly increases with energy over the interval studied. The theoretical values are compatible with the experimental upper bound $k_s \leq 5\times10^{-13}$~cm$^3$~s$^{-1}$. The resemblance of the profiles of the channel-averaged rate constant $k$ (Fig. \ref{results} (c)) and the predictions for the lowest channel (Fig. \ref{results} (e)) suggests that the reaction rate shows a similar dependence on energy across all relevant channels i.e., Ba$^+(6s)$+Rb$(5s)$ Ba$^+(6p)$+Rb$(5s)$ and Ba$^+(6s)$+Rb$(5p)$. 

\section{Summary and conclusions}
\label{discussion}

Cold reactive collisions between laser-cooled Ba$^+$ ions and Rb atoms show features similar to other mixed ion-atom systems such as Ca$^+$+Rb \cite{hall11a, hall13a}, Yb$^+$+Ca \cite{rellergert11a}, Ba$^+$+Ca \cite{sullivan12a} and Yb$^+$+Rb \cite{ratschbacher12a}: a strong dependence of the reaction rate on the electronic state of the reaction partners, the importance of radiative processes  and the formation of molecular ions by RA. Evidence for RA in these systems has first been established in Ref. \cite{hall11a} by mass spectrometry. In the present study, the generation of molecular ions was directly observed by their sympathetic cooling into the Coulomb crystals (see Fig. \ref{scans} (a)). 

As already argued in Ref. \cite{hall11a}, light-assisted processes play an important and often dominant role in the chemistry of ion-atom hybrid systems. In all of the systems studied so far except Yb$^+$+Ca \cite{rellergert11a}, a pronounced acceleration of the rate upon electronic excitation of the reaction partners has been observed. We attribute this observation to the high density of electronic states occurring in mixed ion-atom systems already at moderate excitation resulting in multiple opportunities for RA, RCT and NRCT. Thus, these processes emerge as a common theme in mixed ion-atom systems underlining the rich chemistry that can be observed under the cold conditions achieved in ion-atom hybrid traps. 

On a more general level, hybrid trapping technology has paved the way for the study of ion-atom collisions at an unprecedented level of detail by enabling the determination of state-specific rate constants down to millikelvin collision energies. The exquisite degree of control and sensitivity achieved in these experiments opens up perspectives to gain an improved understanding of ion-atom collision processes that reaches beyond standard pictures such as the classical Langevin model \cite{gao11a, hall12a} and accounts for non-adiabatic, radiative and quantum-collisional effects such as orbiting resonances (see section \ref{rcs} and Refs. \cite{belyaev12a, hall13a}). However, further experimental and theoretical studies are required to characterize these phenomena in more detail.

\section*{Acknowledgements}

We acknowledge support from the Swiss National Science Foundation (grants nr. PP0022\_118921 and PP00P2\_140834 ) and the COST Action MP1001 "Ion Traps for Tomorrow's Applications''. 


\vspace{12pt}



\label{lastpage}

\end{document}